\documentclass{article}

\usepackage{amsmath}
\usepackage{amsfonts}
\usepackage{amssymb}
\usepackage{graphicx}
\usepackage{geometry}
\usepackage{lipsum}
\usepackage{tikz}
\usepackage{subcaption}
\usepackage[color=green!30]{todonotes}

\usepackage{jheppub}

\newcommand{\manifold}{\mathcal M}
\newcommand{\lagrangian}{\mathcal L}
\newcommand{\action}{\mathcal S}
\newcommand{\temperature}{T}
\newcommand{\cc}{\Lambda_C}
\newcommand{\adsradius}{L}
\newcommand{\rtz}{{R^{(3)}(0)}}
\newcommand{\rzz}{{R(0)}}
\newcommand{\horizonradius}{z_\mathrm{h}}
\newcommand{\radiuscutoffscaling}{c}
\newcommand{\hardwallradius}{z_\mathrm{h}}
\newcommand{\boundary}{{ \partial\manifold }}

\newcommand{\pd}[1]{\partial_#1}
\newcommand{\met}{g}
\newcommand{\detg}{\sqrt{-\met}}

\newcommand{\bscalar}{\phi}
\newcommand{\bscalarmag}{|\bscalar|}
\newcommand{\magprof}{\mathit{h}}
\newcommand{\bscalarvac}{|\bscalar_{\mathrm{vac}}|}
\newcommand{\scalarop}{\mathcal{O}_\bscalar}
\newcommand{\vev}{\left\langle\scalarop\right\rangle}
\newcommand{\bsphase}{\Theta}
\newcommand{\scalarmass}{m}
\newcommand{\quarticcoupling}{\lambda}
\newcommand{\pot}{V}

\newcommand{\interactionenergy}{U}
\newcommand{\bmagpert}{v}

\newcommand{\radiuscutoff}{\Lambda}
\newcommand{\ircutoff}{z_\mathrm{cutoff}}
\newcommand{\uvcutoff}{\epsilon}
\newcommand{\potboundary}{{V_\boundary}}

\newcommand{\indicial}{\beta}
\newcommand{\pertindicial}{\alpha}
\newcommand{\constant}{\mathrm{C}}

\title{Holographic Global Vortices with Novel Boundary Conditions}
\date{March 28, 2024}

\author[a]{Markus A.G. Amano,}
\emailAdd{markus.a.amano@gmail.com}

\author[a,b,c]{Minoru Eto}
\emailAdd{meto@sci.kj.yamagata-u.ac.jp}

\affiliation[a]{
  Department of Physics, Yamagata University, Kojirakawa-machi 1-4-12, Yamagata, Yamagata 990-8560, Japan
}

\affiliation[b]{
  Research and Education Center for Natural Sciences, Keio University, 4-1-1 Hiyoshi, Yokohama, Kanagawa 223-8521, Japan
}

\affiliation[c]{
  International Institute for Sustainability with Knotted Chiral Meta Matter(SKCM$^2$), Hiroshima University, 1-3-2 Kagamiyama, Higashi-Hiroshima, Hiroshima 739-8511, Japan
}

\begin{document}

\begin{flushright} 
  YGHP-24-03
\end{flushright} 

\abstract{
  The AdS/CFT correspondence has significantly impacted the study of strongly coupled systems, providing insights into various condensed matter phenomena through its holographic duality. 
  This paper introduces an alternative approach to the breaking of the global $U(1)$ symmetry in the bulk in two asymptotically AdS spacetimes: AdS plus hard wall and AdS Blackbrane. 
  We explore a $(3+1)$-dimensional bulk $U(1)$ symmetry-breaking phase vacuum within a global $U(1)$ $\phi^4$ field theory, without the gauge field present in prior models. 
  We find the symmetry-breaking vacuum requires that the mass squared is proportional to the quartic coupling. 
  We also investigate numerical solutions of topologically stable vortex strings extending into the bulk.
  We find evidence that the full UV expansion is dual to a point-like boundary excitation.
}

\maketitle

\section{Introduction}

The advent of the Anti-de Sitter/Conformal Field Theory (AdS/CFT) correspondence, as introduced by Maldacena \cite{Maldacena:1997re}, revolutionized our understanding of strongly coupled systems. 
This theoretical construct bridges string theory in negatively curved AdS spacetimes with CFTs that inhabit the spacetime boundary\footnote{
  Here, the term "boundary" refers to the conformal boundary of the bulk space, distinct from the null boundary of Minkowski space. In this context, signals within the bulk can reach the boundary and return in a finite time frame.
}.
Throughout the past two decades, the AdS/CFT correspondence has been applied extensively within nuclear physics and condensed matter, among other fields.

In the field of condensed matter physics, a rich spectrum of phenomena has been modeled using the AdS/CFT framework, revealing gravitational duals for each. 
Prominent examples include the quantum Hall effect \cite{Hartnoll:2007ai, Fujita:2009kw}, the Nernst effect \cite{Hartnoll:2007ih}, superconductivity \cite{Hartnoll:2008kx}, and Lifshitz scaling \cite{Kachru:2008yh}.

The foundational work of Hartnoll et al. 
\cite{Hartnoll:2008kx, Hartnoll:2008vx} laid the groundwork for holographic studies of superfluidity and superconductivity. 
Their research demonstrated that, below a certain critical temperature, a scalar field, which initially has a negligible expectation value, develops a non-zero value, heralding the emergence of spontaneous symmetry-breaking of a dual $U(1)$ symmetry. 
They particularly noted a sharp increase in direct current (DC) conductivity at the critical temperature, characteristic of a holographic superconductor.

Typically, the $(3+1)$-dimensional Abelian-Higgs model serves as the bulk theory for such investigations. 
There is also much attention to the understanding holographic dual of strongly coupled Bose-Einstein Condensates (BECs).
Regarding BECs, various studies have simulated the dynamics of strongly coupled $U(1)$ physics with inhomogeneous phases \cite{Keranen:2009ss, Keranen:2009re}, with multiple component scalars \cite{Yang:2020wes, Yang:2019ibe, Yang:2022foe, Kasamatsu:2003ktu}, and under rotation \cite{Sonner:2009fk, Gregory:2014uca, Srivastav:2023qof}.
Further studies have expanded on these condensed matter phenomena of vortices \cite{Dias:2013bwa, Albash:2009iq, Wittmer:2020mnm, Wittmer:2021oxf} and lattice phases \cite{Su:2023vqa, Xia:2019eje}.
A central aspect in this line of research is the breaking of $U(1)$ symmetry within certain boundary theories. 
This symmetry breaking, often facilitated by a non-zero $U(1)$ chemical potential, dual to a gauge field in the bulk.
The $U(1)$ breaks at a critical temperature, resulting in a scalar condensate.

In this work, we present an alternative approach to the breaking of the global $U(1)$ symmetry in the bulk. 
Our model differs from previous works by the presence of a bulk $U(1)$ breaking scalar vacuum.
We find that this vacuum solution's existence requires that the scalar's mass squared, $\scalarmass^2$, becomes negative, proportional to the quartic coupling, $\quarticcoupling$, i.e. $\quarticcoupling = -\scalarmass^2\adsradius^2$, where $\adsradius$ is the radius of AdS.
We will also probe two bulk AdS geometries: AdS with a hard wall and AdS Black Brane.
We find that this scalar vacuum is stable against sourceless perturbations.

Our analysis explores topologically stable one-dimensional bulk vortices. 
We examine single vortex and pair vortex configurations.
For the pair of vortices we simplify the physical model by only considering well-separated vortex strings.
We find evidence that the vortices are dual to a boundary field state which has a localized non-zero scalar vacuum expectation value (VEV).
The candidate VEV condensates are found to center around vortex termination points.
However, we find that vortices severely warp the geometry at the boundary; bulk vortices up to the boundary cannot be treated as probe fields.
This possibly indicates that bulk vortices can be treated as nucleation sites for a phase transition and required us to impose a UV-cutoff in the bulk coordinate.

Furthermore, we find numerical radial profiles of the vortex string pairs.
We find that vortex-vortex and vortex-antivortex pairs repel and attract, respectively.
Furthermore, we find a possible link with the near boundary behavior of radial profile and the scalar VEV.
We also find that these vortices can have valid radial profiles only above a certain minimum temperature.

This paper is organized in the following way.
In Section \ref{sec:analysis}, we will introduce the bulk theory and derive the $U(1)$ symmetry breaking vacuum.
We will also will analyze the single vortex equation of motion for a vortex centered at $x=y=0$ that ends on boundary.
After, we will analyze the radial profiles for well-separated vortex strings.
In Section \ref{sec:results}, we will cover analytical and numerical results about the vortex strings.
Finally, we will conclude in Section \ref{sec:conclusion}

\section{Analysis}
\label{sec:analysis}

Our bulk gravitational theory is global $\phi^4$ $U(1)$ scalar plus Einstein Gravity with a negative cosmological constant.
The background geometry is taken to be asymptotically AdS.
The bulk dimensions are $(3+1)$D.
The action takes the simple form of \eqref{eq:scalargravityaction}.
\begin{equation}
  \label{eq:scalargravityaction}
  \begin{aligned}
    \action = \action_\mathrm{gravity} + \action_{\mathrm{matter}} = &\int \detg \left(R - 2\cc \right) \\
    - &\int \detg \left(\met^{\mu\nu}\left(\pd{\mu}\bscalar\right)\left(\pd{\nu}\bscalar\right)^\star + \pot(|\bscalar|^2)
    \right)
    \end{aligned}
  \end{equation}
  \begin{equation}
    \label{eq:bscalareom}
    -\frac 1\detg \pd\mu\left(\met^{\mu\nu}\detg\pd\nu\bscalar\right) + \pot'(|\bscalar|^2)\bscalar = 0
  \end{equation}
  $\bscalar$ is the complex scalar field.
  $\pot$ is the quartic potential such that 
  \begin{equation}
    \label{eq:potential}
    \pot(|\bscalar|^2) = \frac \quarticcoupling2 \left( |\bscalar|^2 \right)^2 + \scalarmass^2 |\bscalar|^2
    = \frac \quarticcoupling2 \left( |\bscalar|^2 + \frac{\scalarmass^2}{\quarticcoupling} \right)^2 - \frac{\left(\scalarmass^2 \right)^2}{2 \quarticcoupling}\,.
  \end{equation}
  There are several choices we could make for the analysis in this paper like in terms of deforming the geometry.
  For this work we will restrict ourselves to consider a simple, non-deformed asymptotically AdS geometry in which has a metric of the form \eqref{eq:metric} with probe matter fields.
  \begin{equation}
    \label{eq:metric}
    \met = \frac {L^2}{z^2} \left( -f(z) dt^2 + \frac 1{f(z)}dz^2 + \delta_2 \right)
  \end{equation}
  $\delta_2 \equiv dx^2 + dy^2$ is the 2D Euclidean metric. $t,\,x,\,y \in \mathbb R$, and $z \in \mathbb R^+$.
  Furthermore, $f(z)$ is the blackening factor.
  We consider two cases for $f(z)$.
  The first case covered is $f(z) = 1$, AdS.
  For AdS, we will artificially introduce a hard wall (HW).
  "hard wall" here means a surface were the scalar fields are set by hand to end. 
  The hard wall is a surface at points were $z = \hardwallradius$.
  The second case is $f(z) = 1 - z^3/\horizonradius^3$, AdS black brane; $\horizonradius$ is the horizon radius.
  We find in Section \ref{sec:boundary_conditions}, the radial profiles of the vortex strings must obey Neumann boundary conditions on the horizon (and on the boundary).
  That is to say, that vortices end on $\horizonradius$ and start on the boundary.

  The keen-eyed reader might have observed that the hard wall radius and the horizon radius are denoted by the same symbol. 
  Since they both function in the same way to terminate the vortices and lead to similar vortex profiles, we will keep the notation for simplicity.

  \subsection{Near Boundary Expansion}
  \label{sec:near_boundary_expansion}

  We would like to find vacuum solutions to \eqref{eq:bscalareom} near conformal boundary such that $\bscalar \sim z^0$.
  Assuming a real homogeneous solution, $\left(\bscalar =  z^\indicial/\adsradius^{\indicial+1} \right)$ near the boundary (so $f(z) \approx 1$) \eqref{eq:indicial_scalar}.
  \begin{equation}
    \label{eq:indicial_scalar}
    - \indicial \left(\indicial - 3\right) \adsradius^{-\beta-3} z^\indicial + m^2 \adsradius^{-\beta-1}z^\indicial + \adsradius^{-3\beta-3}\quarticcoupling z^{ 3\indicial } = 0
  \end{equation}
  For $z \to 0$, $\indicial \left(\indicial - 3\right) = \scalarmass^2 \adsradius^2$ is asymptotically true if $\indicial > 0$.
  Nonetheless, in this paper, we take another admissible asymptotic solution.
  In this solution $\beta = 0$.
  Namely, we take $\bscalar \sim z^0$, so that
  \begin{equation}
    \adsradius^2\scalarmass^2 = - \quarticcoupling\,.
  \end{equation}
  This latter solution implies that the $U(1)$ vacuum in the bulk preserves the AdS asymptotics as long as $\adsradius^2\scalarmass^2 = - \quarticcoupling$.
  Further, there is a $U(1)$ symmetry-breaking phase.
  The vacuum value is $\bscalarvac^2 = -\scalarmass^2/\quarticcoupling$ which conveniently simplifies to $\bscalarvac^2 = 1/\adsradius^2$.
  We can absorb $\adsradius$ dependence by the following rescaling of our fields as
  $\met_{\mu\nu}\rightarrow \met_{\mu\nu}/\adsradius^2$,
  $\bscalar\rightarrow \bscalar \adsradius$,
  $\scalarmass\rightarrow \scalarmass/\adsradius$.
  Nevertheless, there is a drawback to this "constant" alternative boundary condition.
  The departure from the standard boundary conditions thus requires more care in whether or not the action/energies generated by such vacuum is renormalizable.
  The energy momentum tensor is a linear combination of Lagrangian and square of "$\pd{\mu}$".
  The solution is constant so all gradient terms vanish.
  The source of any divergence would then have to come from the potential term in the Lagrangian.
  In the $\bscalarvac^2 = 1$ vacuum, the current potential is a constant $-\quarticcoupling/2$.
  Instead, we may consider the new non-negative potential, shifted up by $\quarticcoupling/2$, since this alteration leaves the equation of motion unchanged. 
  Now the potential is zero in this vacuum, leaving no contributions to the energy stress tensor.
  This again allows us to treat the $\bscalar$ as a probe field.
  From this point onwards, the potential \eqref{eq:potential} can be written as
  \begin{equation}
    \pot(\bscalarmag^2) = \frac \quarticcoupling2 \left( \bscalarmag^2 - 1\right)^2\,.
  \end{equation}

  To test the linear stability of this vacuum we introduce perturbation of $\bscalar$.
  Perturbing around this vacuum, we perturb $\bscalar$ as \eqref{eq:perturbedscalar}.
  We assume that $|\bmagpert| \ll |\bscalar_{\mathrm{vac}}| \equiv 1$ and $\eta \ll 1$.
  \begin{equation}
    \label{eq:perturbedscalar}
    \bscalar =  1 + \bmagpert + i \eta
  \end{equation}
  \eqref{eq:perturbedscalar} can be plugged into the equations of motion and further expanded around small $\bmagpert$ near the boundary.
  For a homogeneous field, the near boundary expansion will have a leading behavior of $\bscalar \rightarrow  1 + \bmagpert^{(0)} z^\pertindicial + i\eta^{(0)} z^\gamma $.
  The following equation gives an indicial equation similar to a massive scalar field with a mass squared of $2\quarticcoupling$.
  Also associated with the symmetry-breaking phase, the $\eta$ is a massless scalar field.
  Such equivalent scalar perturbations are known to be linearly stable \cite{Starinets:2002br, Nunez:2003eq}.
  \begin{equation}
    \label{eq:pert_indicial_scalar}
    \begin{aligned}
      \left( - \pertindicial \left(\pertindicial - 3\right)  + 2 \quarticcoupling \right)z^\pertindicial \bmagpert^{(0)} + \mathrm O\left(\left(\bmagpert^{(0)}\right)^2\right) &= 0\\
      - \gamma \left(\gamma - 3\right)z^\gamma \eta^{(0)} + \mathrm O\left(\left(\eta^{(0)}\right)^2\right) &= 0
    \end{aligned}
  \end{equation}
  \eqref{eq:pert_indicial_scalar} implies two solutions for $\pertindicial$, 
  \begin{equation}
    \label{eq:indicial_equation}
    \pertindicial = 3/2 \pm \sqrt{9/4 + 2\lambda} \,.
  \end{equation}
  Since $\quarticcoupling > 0$, one of the solutions has diverging behavior for $\pertindicial$, such that $\pertindicial < 0$.
  This boundary behavior would violate a earlier assumption that $|\bmagpert| \ll |\bscalar_{\mathrm{vac}}| = 1$.
  So in order to respect this constraint, we need to take the positive solution in \eqref{eq:indicial_equation}.
  The resulting perturbation scalar field must not source any current on the boundary.
  $\eta$ does not have this restriction because both of its Frobenius indicial exponents {($0$ or $3$)} are non-negative.
  Now focusing on the scalar portion of the action \eqref{eq:scalargravityaction}, it can now be written as 
  \begin{equation}
    \label{eq:scalaraction}
    \action_{\mathrm{scalar}} = - \int \detg \left( \met^{\mu\nu}\left(\pd{\mu}\bscalar\right)\left(\pd{\nu}\bscalar\right)^\star 
      + \frac \quarticcoupling2 \left( |\bscalar|^2 - 1\right)^2
    \right)\,.
  \end{equation}
  %
  %

  \subsection{Single Vortex}
  \label{sec:single_vortex}

\begin{figure}[htbp]
    \centering
    \begin{minipage}{.45\textwidth}
      \centering
      \includegraphics[width=0.95\textwidth]{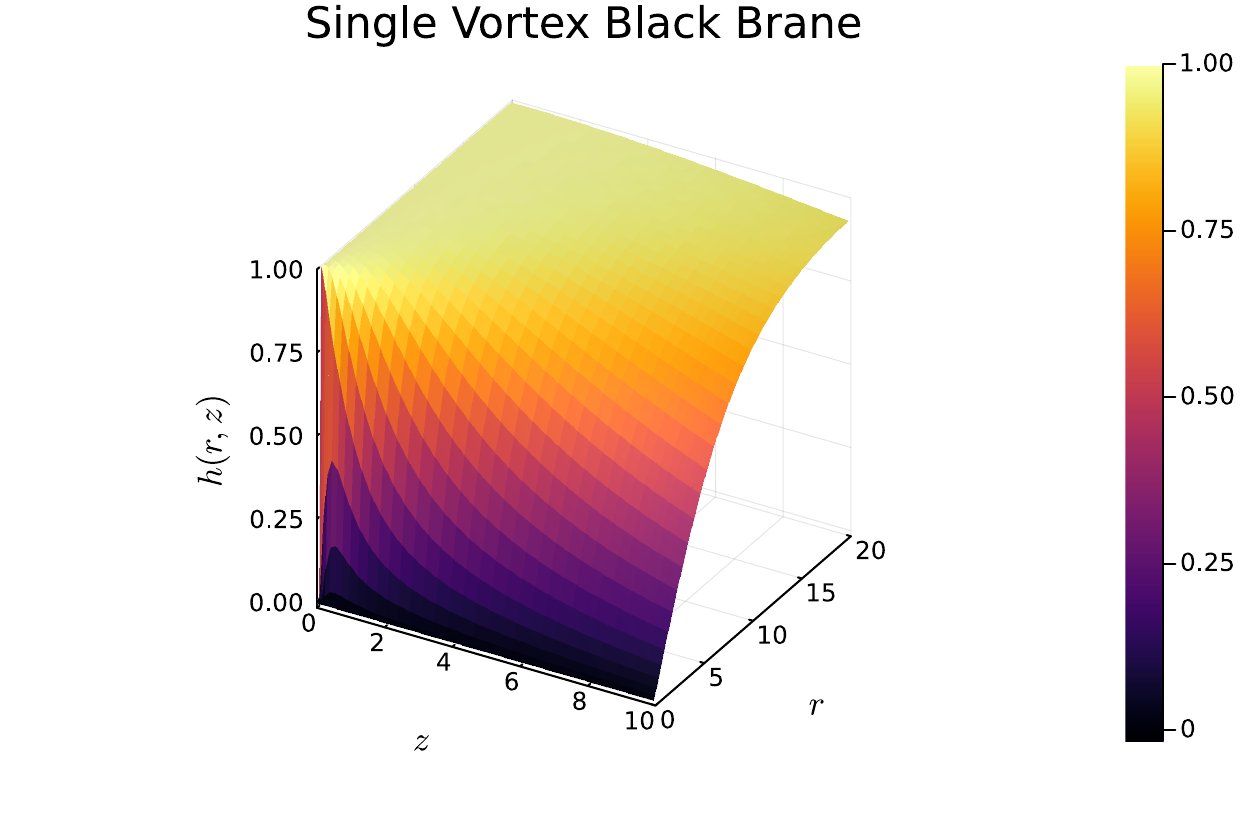}
    \end{minipage}%
    \hspace{0.05\textwidth}
    \begin{minipage}{.45\textwidth}
      \centering
      \includegraphics[width=0.95\textwidth]{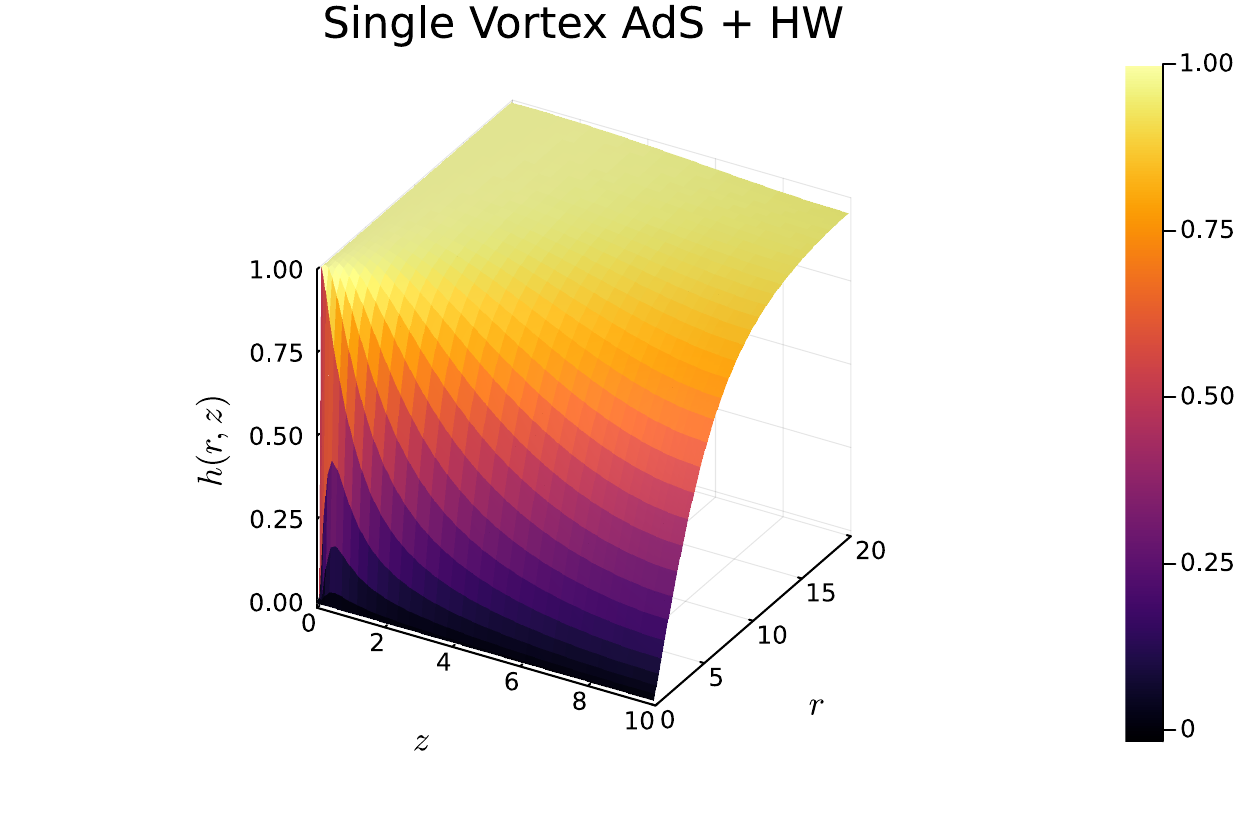}
    \end{minipage}
    \caption{Vortex $\magprof$ for the AdS with hard wall (left) and black brane (right) case where $\quarticcoupling = 1$ and $\hardwallradius = 10$.}
    \label{fig:single_vortex_surface_plots}
  \end{figure}

\begin{figure}[htbp]
    \centering
    \begin{minipage}{.45\textwidth}
      \centering
      \includegraphics[width=\textwidth]{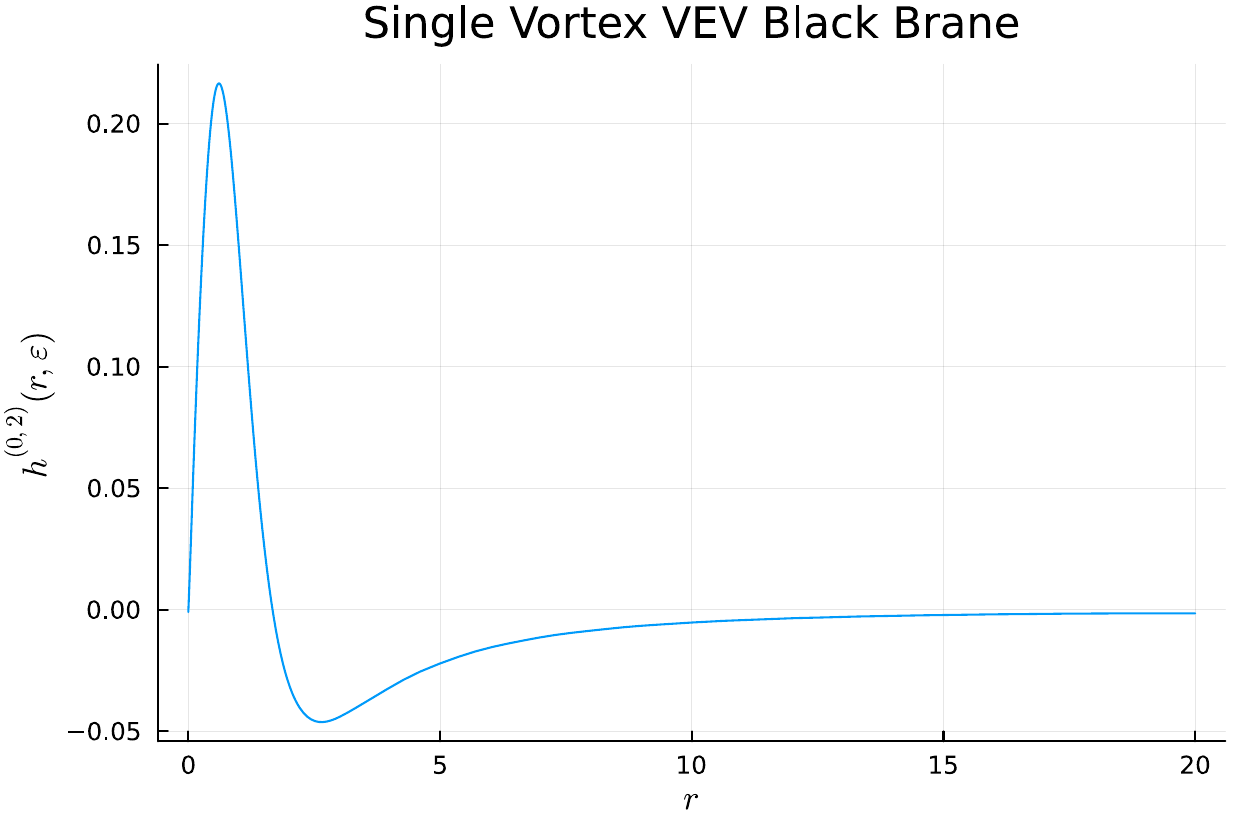}
    \end{minipage}%
    \hspace{0.05\textwidth}
    \begin{minipage}{.45\textwidth}
      \centering
      \includegraphics[width=\textwidth]{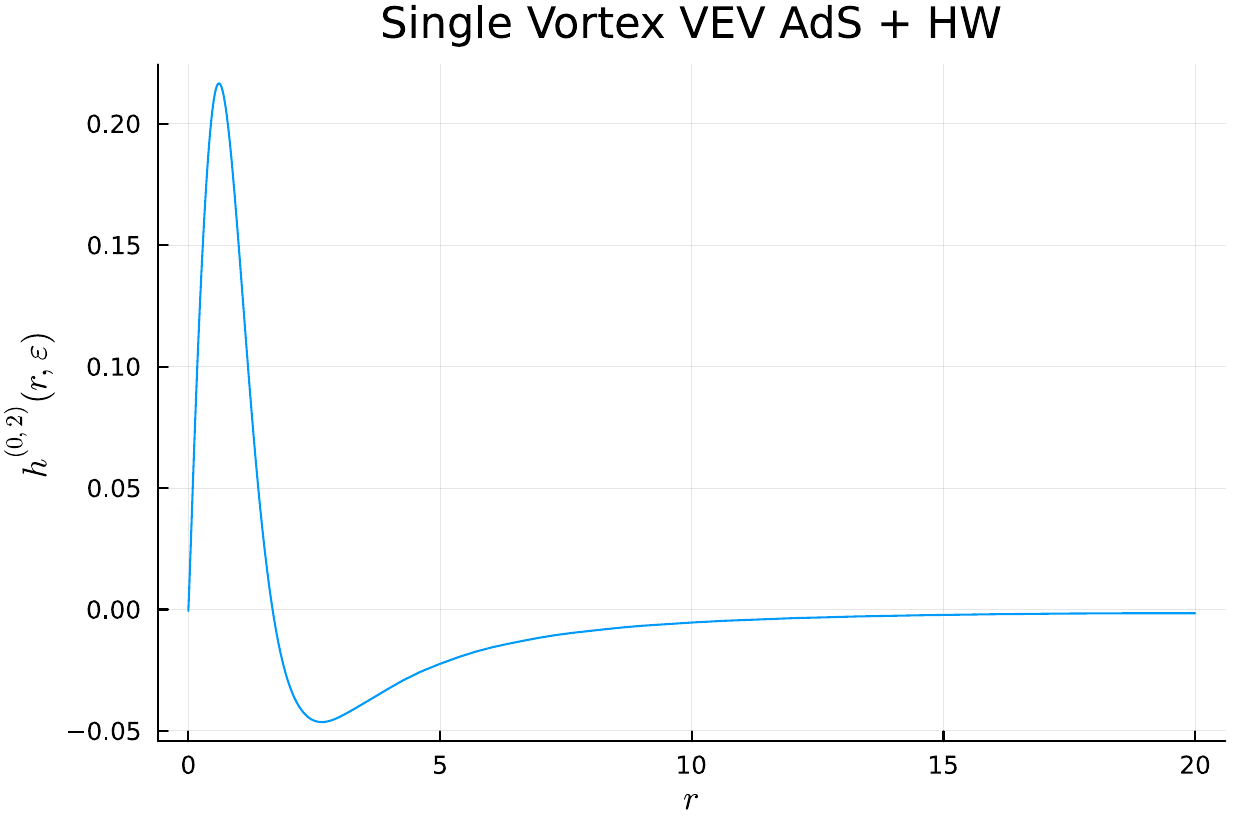}
    \end{minipage}
    \caption{ Evaluated with an $\uvcutoff \approx 1.46$, these are plots of the second derivative of $\magprof$ with respect to $z$. Plots for the black brane and for the AdS + HW case indicate the a tapering VEV for large lateral distances from the core of the region where the bulk vortex approaches the boundary.}
    \label{fig:single_vortex_surface_vev_plots}
  \end{figure}

  In this section we calculate the equation of motion for a single vortex solution.
  Centering a vortex string at $x = y = 0$ parallel to the $z$ axis, we take the following ansatz.
  \begin{equation}
    \label{eq:singlevortansatz}
    \bscalar = \magprof(r, z) e^{i k\theta}
  \end{equation}
  where $k \in \mathbb Z$.
  Plugging in \eqref{eq:singlevortansatz} into \eqref{eq:bscalareom}, we get the following PDE.
  \begin{equation}
    -\frac 1\detg \pd\mu\left(\met^{\mu\nu}\detg\pd\nu\bscalar\right) + \pot'(|\bscalar|^2)\bscalar = 0
  \end{equation}
  %
  \begin{equation}
    \label{eq:singlevortex_mageom}
    z^4 \pd z\left(\frac{f}{z^2}\pd z \magprof\right)
    +\frac { z^2 }{r} \pd r\left(r\pd r \magprof\right)
    -\frac {z^2}{r^2} k^2 \magprof
    - \quarticcoupling \left( \magprof^2 - 1\right) \magprof 
    = 0
  \end{equation}
  \eqref{eq:singlevortex_mageom} is a second order non-linear boundary value problem.
  We enforce boundary conditions to make the solution to \eqref{eq:singlevortex_mageom} well-defined.
  At $r = 0$ we set the Dirichlet $\magprof(0, z) = 0$.
  For $r = \infty$ we set a Dirichlet boundary condition $\magprof(\infty, z) = 1$.
  At both ends of the $z$ domain we set Neumann boundary conditions where $\magprof^{(0, 1)}(r, 0) = \magprof^{(0, 1)}(r, \horizonradius) = 0$.
  One might think that a separation of variables could be applied here, but geometrical factor $z^2$ obstructs such a solution.
  Instead of analytical solutions we find a family of numerical solutions.
  For an example solution we solved for $\quarticcoupling = 1$ and $\horizonradius = 10$ for both the AdS+HW and black brane case Figure \ref{fig:single_vortex_surface_plots}.
  \begin{figure}[htbp]
    \centering
    \begin{minipage}{.45\textwidth}
      \centering
      \includegraphics[width=\textwidth]{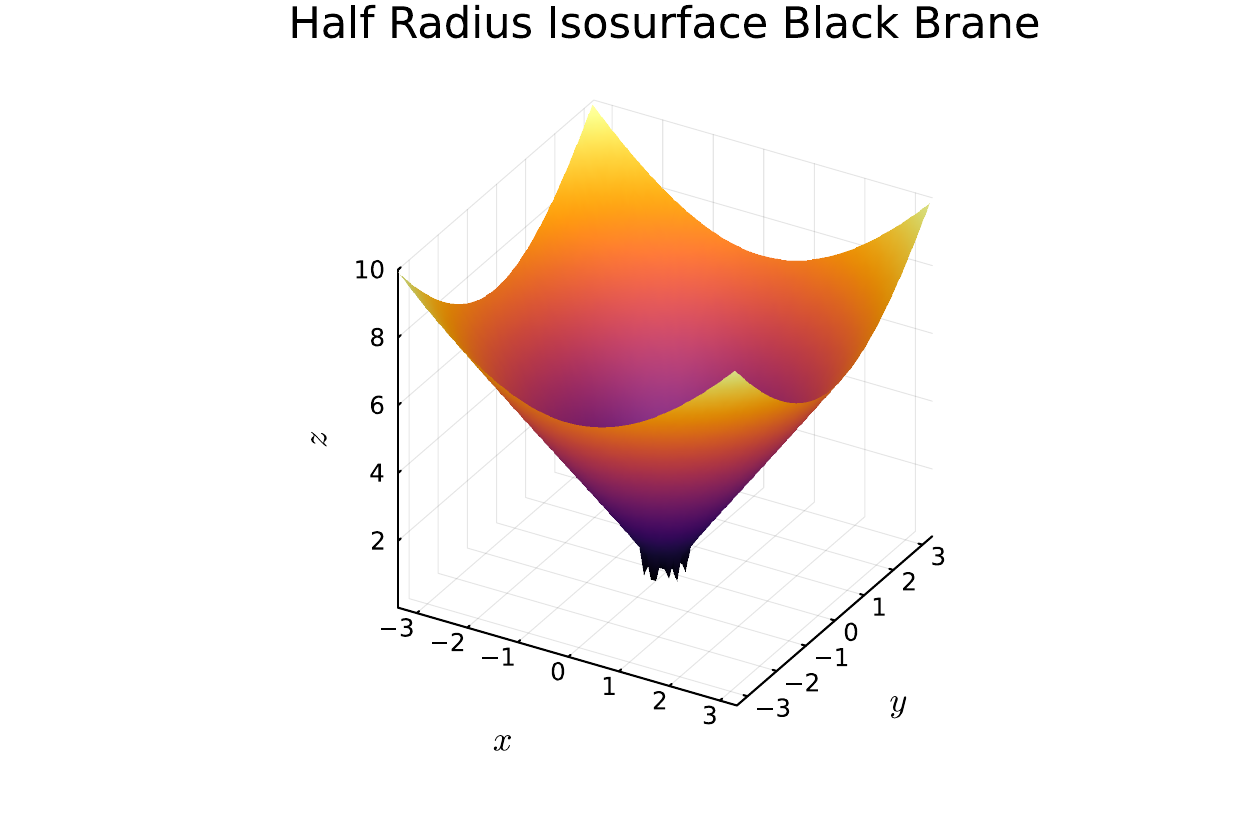}
    \end{minipage}%
    \hspace{0.05\textwidth}
    \begin{minipage}{.45\textwidth}
      \centering
      \includegraphics[width=\textwidth]{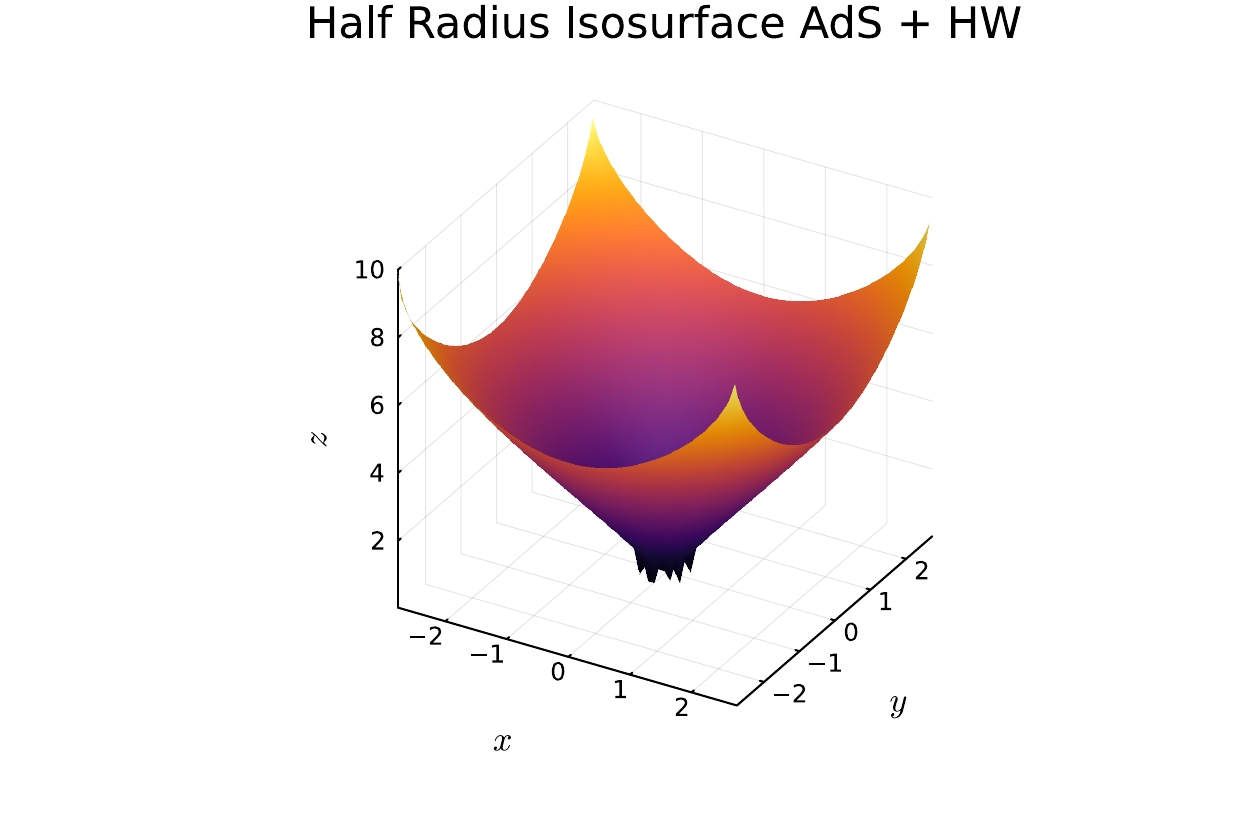}
    \end{minipage}
    \caption{
      These are surface plot where $\magprof(r, z) = 1/2$.
      The black brane (left) is more similar to a cone shape than the AdS plus hard wall which has a more curvature apparently.
      Both plots where calculated with a numerical solution where $\quarticcoupling = 1$ and $\hardwallradius = 10$.
    }
    \label{fig:single_vortex_surface_cone_plots}
  \end{figure}
  Even though, we found numerical solutions, they are similar to vortices of different $\quarticcoupling$ at different $z$ cross-sections.
  The width of the vortex is proportional to $z/\sqrt{\quarticcoupling}$.
  In Figure \ref{fig:single_vortex_surface_cone_plots}, the dependence of vortex string width to bulk coordinate can be seen.
  In the other direction, for near the boundary the vortex's width shrink as if under a large $\quarticcoupling$ to a cross-sectionally a point vortex (up to the resolution of the numerical method).
  It appears that at the boundary the vortex solution is singular.
  It is still not clear, what boundary state corresponds to static single vortex solutions.
  Usually, the VEV of the scalar operator, $\vev$, is related to subleading behavior of the scalar field.
  For the free scalar field case, the $\vev$ is proportional to the subleading behavior of on-shell solution near the boundary.
  For a generic on-shell scalar field, $\phi$, this looks like {$\phi \sim \phi^{(0)} z^{ \Delta_- } + \constant_1 \vev z^{ \Delta_+ }$}.
  $\Delta_\pm$ and $\constant_1$ are indicial exponents and a constant respectively.
  A proxy of the scalar VEV, $\vev$, is the value of the on-shell bulk field on the boundary.
  Since the first derivative is set to zero via a Neumann boundary condition, we can use the second derivative as a proxy for $\vev$.
  This is plotted in Figure \ref{fig:single_vortex_surface_vev_plots}.
  The plots suggest that a dual vortex is a vacuum disturbance in the boundary theory.
  Possibly, in the full UV theory, vortices correspond to point nucleation sites on the boundary.
  Such a finding corroborates with Section \ref{sec:near_boundary_expansion}.
  %

  To continue in this work, we will impose a cutoff around the center of vortices on the boundary where $0 < \uvcutoff^2 < r^2 + z^2$.
  Furthermore, we are dealing with global vortices, so the tension/mass of the vortex has a log divergence.
  \footnote{
    Interestingly enough for the black brane for a gauged vortex profile $h \approx \tanh(r/z^2)$, we find the finite part of the action is proportional to $\int dt M/\temperature$, where $M$ is the mass of the vortex.
  }
  Since the width of the vortex is finite, we can treat them as vortex string approximations with enough spacing in the transverse direction.

  \subsection{Vortex String Approximation}
  \label{sec:vortexapproximation}

\begin{figure}[htbp]
    \centering
    \includegraphics[width=\textwidth]{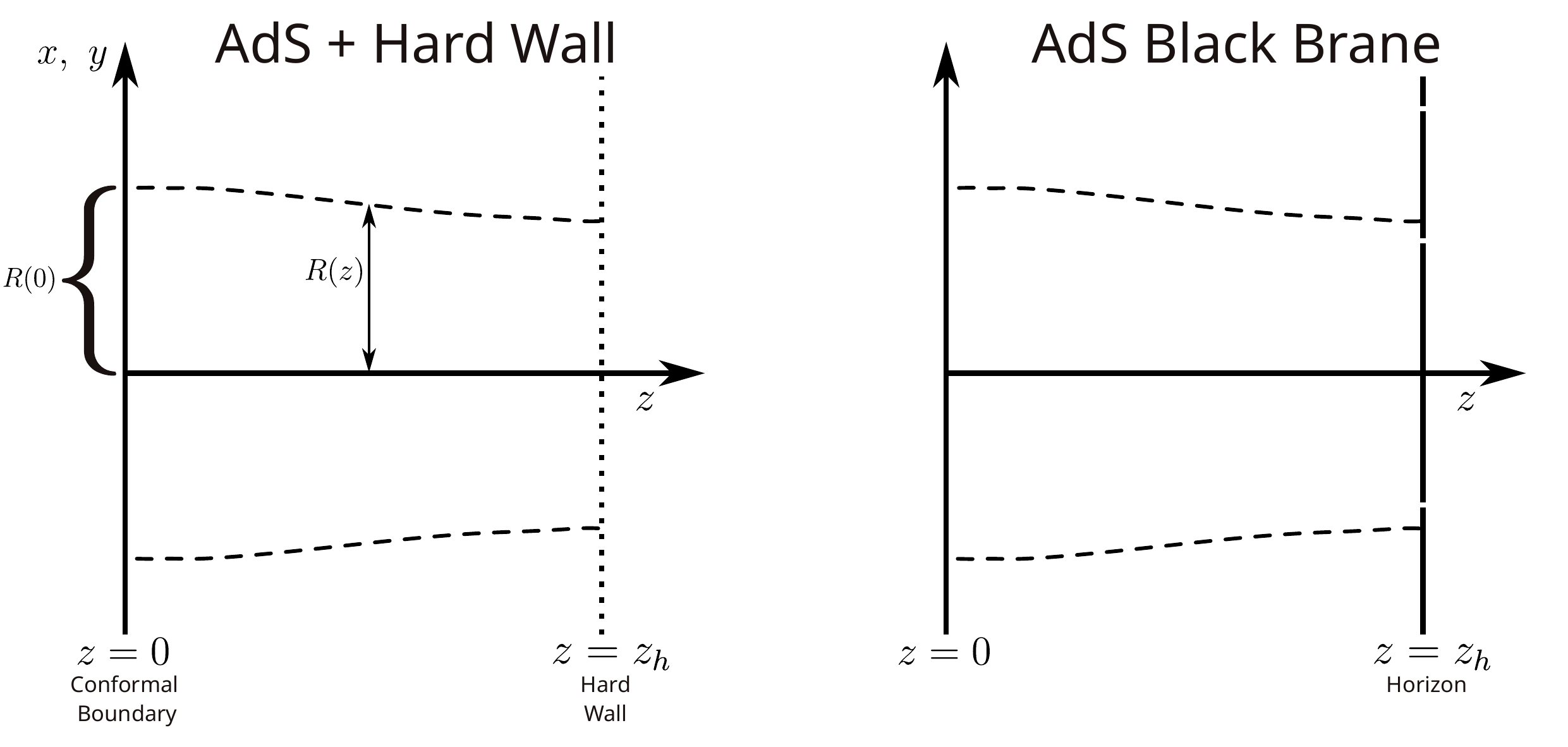}
    \caption{
      Here are two sketches of vortex strings (as dashed lines) in two different geometries.
      The left diagram is anti-de Sitter with an hard wall.
      The vortex string (and the scalar) field is artificially set to end at $z = \hardwallradius$.
      The right diagram is the anti-de Sitter black brane case.
      Here, the vortex string like before ends on the black brane's horizon at $z = \horizonradius$.
      The field theory side of the duality is physics that takes place at the spatially $2$D conformal boundary, $z=0$.
    }
    \label{fig:radial_profile_diagram}
  \end{figure}

  For a sufficiently large separation relative to the size of the vortices, point-like approximations can be justified for a pair of vortices. 
  In the context of a higher-dimensional framework, such an object can be conceptualized as a string that penetrates the bulk, originating from the boundary. 
  This is illustrated in Figure \ref{fig:radial_profile_diagram}.
  This representation allows for a simplified mathematical description, where a single vortex string can be approximately described by the following equation:
  \begin{equation}
    \label{eq:vortex_ansatz}
    \bscalar_i = e^{ \pm i \bsphase_i}
  \end{equation}
  where, $\bsphase_i$ denotes the angle measured around a point lateral to a given holographic position, $z$. 
  For a string, the central location from which $\bsphase_i$ is measured will typically depend on $z$. 
  In the case of vortex-vortex or vortex-antivortex interactions, a product of two instances of \eqref{eq:vortex_ansatz}, representing two distinct locations, serves as an approximation.
  Given the translational and rotational symmetries along the transverse directions, it is practical to position the vortices symmetrically around the origin and along the $x$-axis, as detailed in \eqref{eq:vortexapproximation}-\eqref{eq:bsphaseparam2}.
  \begin{equation}
    \bsphase_{12} = \bsphase_1 + s\bsphase_2
    \label{eq:vortexapproximation}
  \end{equation}
  \begin{align}
    \label{eq:bsphaseparam1}
    \tan( \bsphase_1 ) &= \frac{y}{x - R(z)}\\
    \label{eq:bsphaseparam2}
    \tan( \bsphase_2 ) &= \frac{y}{x + R(z)}
  \end{align}
  \eqref{eq:vortexapproximation} can thus be interpreted as the combined angle measured from two distinct vortices located symmetrically at positions $(x, y) = (\pm R(z), 0)$.
  $s = +1$ for vortex-vortex, and $s=-1$ for vortex-antivortex.

  \section{Results}
  \label{sec:results}

  This section is dedicated to presenting the numerical and analytical results derived from the methodologies and theories discussed in the previous section. 

  \subsection{Radial Profile Analysis}
  \label{sec:radialprofileanalysis}

  We would like to examine the interaction energy between two approximated string-like vortices.
  To approximate this energy, we take the energy of the full approximation with ${\bscalar_{ 12 } := \exp(i(\bsphase_1 + s\bsphase_2))}$ \eqref{eq:vortexapproximation} and subtract the energy contributions of ${ \bscalar_1 := \exp(i\bsphase_1) }$ \eqref{eq:bsphaseparam1} and ${ \bscalar_2 := \exp(is\bsphase_2) }$ \eqref{eq:bsphaseparam2} where $s = \pm 1$.
  These vortex strings have an interaction potential as described by \eqref{eq:A0vortexpotetial}.
  In terms of Lagrangians, this takes the form of \eqref{eq:interactionenergy}.
  \begin{equation}
    \label{eq:interactionenergy}
    \interactionenergy := -\int dx^3 ( \lagrangian(\bscalar_{12}) - \lagrangian(\bscalar_2) - \lagrangian(\bscalar_1) )
  \end{equation}
  Using the vortex string ansatz and the scalar Lagrangian \eqref{eq:scalaraction}, \eqref{eq:interactionenergy} can be written as.
  \begin{equation}
    \label{eq:A0vortexpotetial}
    \interactionenergy = s\int dz\, dx^2 \frac{1}{2} \sqrt{g} g^{ij}( \pd{i}\bsphase_1 )( \pd{j}\bsphase_2 )\,.
  \end{equation}
  The vortices are parameterized by \eqref{eq:bsphaseparam1} and \eqref{eq:bsphaseparam2}\footnote{ i,j,k,\ldots are boundary coordinate indices.}.
  After integrating out the $x$ and $y$ directions in \eqref{eq:A0vortexpotetial}, we obtain \eqref{eq:A0vortexpotetialintz}.
  \begin{equation}
    \label{eq:A0vortexpotetialintz}
    \interactionenergy(2\rzz ) = \int dz \frac { \pi  s }{z^2} \left(2 \potboundary(R)-f(z) R'(z)^2 \left(\potboundary(R)+1\right)\right)
  \end{equation}
  where
  \begin{math}
    \potboundary(R) := \log \left(\left(\radiuscutoff^2/R^2 + 1\right)/4\right).
  \end{math}
  $\potboundary(R)$ is the well known potential between point vortices in $(2+1)$D \cite{Smith:2008xx, Shellard:1987bv, Nakano:2008nas, Ovchinnikov:1998tge}.
  $\radiuscutoff$ is an IR-cutoff in the boundary radius direction, $r$.
  This is needed since for global vortices, because there is a log divergence.
  Note that $R$ is a function of $z$.
  To find the extremized energy contributing configuration, we seek $R$ such that \eqref{eq:A0vortexpotetialintz} is extremized.
  The resulting Euler-Lagrange equation is \eqref{eq:vertStaticEom}.
  \begin{equation}
    \label{eq:vertStaticEom}
    \begin{aligned}
    &-R'(z) \Big(R(z) \big(2 f(z)-z f'(z)\big) \big(R(z)^2+\radiuscutoff^2\big)
    \times\big(\potboundary(R)+1\big)+\radiuscutoff^2 z f(z) R'(z)\Big)\\
    &+z f(z) R(z) \big(R(z)^2+\radiuscutoff^2\big) R''(z) \left(\potboundary(R)+1\right)
    =2 \radiuscutoff^2 z
    \end{aligned}
  \end{equation}
  %
  %
  We find in Section \ref{sec:boundary_conditions} that $R$ must obey two Neumann boundary conditions:
  \begin{equation}
    \label{eq:vertStaticBC}
    R'(0) = R'(\horizonradius) = 0\,.
  \end{equation}

  \subsection{Neumann Boundary Conditions}
  \label{sec:boundary_conditions}
  We would like to understand the behavior of our $R$ fields near the boundary and the horizon/hard wall.
  We first will expand $R$ around the boundary, $z=0$, with a Taylor series \eqref{eq:bsRboundaryexpansion}, solving \eqref{eq:vertStaticEom} order by order. 
  \begin{equation}
    \label{eq:bsRboundaryexpansion}
    R(z) = \sum_i \frac 1{i!} R^{(i)}(0) z^i
  \end{equation}
  where $R^{(i)}(z) \equiv d^iR(z)/{dz}^i$.
  We plug \eqref{eq:bsRboundaryexpansion} into \eqref{eq:vertStaticEom} to solve it order by order in $z$ with $R^{(i)}(0)$.
  At the leading order, we find $R^{(1)}(0) = 0$. 
  This can be verified by the reader by looking at Appendix \ref{app:eom_expansion}.
  At the next leading order $R^{(2)}(0)$ can be solved in terms of $\rzz$.
  However, $\rtz$ cannot be determined by lower order coefficients. 
  In later sections, both $\rzz$ and $\rtz$ can be determined to find solutions to the boundary value problem, \eqref{eq:vertStaticEom}.

  We also perform a near horizon analysis for the black brane background.
  We perform a Taylor series around $z=\horizonradius$ \eqref{eq:bsRhorizonexpansion}. 
  Again, we solve the equations of motion order by order in $z$ with \eqref{eq:bsRhorizonexpansion}.
  \begin{equation}
    \label{eq:bsRhorizonexpansion}
    R(z) = \sum_i^\infty \frac 1{i!} R^{(i)}(0) (z - \horizonradius)^i
  \end{equation}
  We find $R^{(1)}(\horizonradius) = 0$ at leading order.
  Analogously, we also enforce a Neumann boundary condition on the hard wall case at $z = \hardwallradius$.

  Regardless of background, these two Neumann boundary conditions along with \eqref{eq:vertStaticEom} admit constant $R$ solutions for large $\horizonradius$.
  In either background, a shooting method can be employed to solve this boundary value problem.
  In short, the shooting method (more in Appendix \ref{app:shooting}) involves setting $\rzz$ on the boundary and an initial $\rtz$; we use a root finding algorithm to find $\rtz$, such that both Neumann boundary conditions are satisfied.
  It is apparent that in the general case, the stated boundary value problem is intractable analytically for the authors writing this paper.
  In lieu of general analytical results, numerical results are presented in Section \ref{sec:numerical_results}.

  \paragraph{Interaction Energy Integration} 
  %
  \eqref{eq:A0vortexpotetialintz} is generically divergent.
  This is due to the log term but nevertheless was to be expected because the single vortex has a divergent on-shell action.
  To deal with this divergence, we need to expand UV cutoff such that $z > \uvcutoff > 0$ and also $(x\pm R)^2 + y^2 > \uvcutoff > 0$ similar to Section \ref{sec:single_vortex}.
  This UV cutoff, $\uvcutoff$, and IR cutoff, $\radiuscutoff$ both limit the length scale of these vortex strings.

  \paragraph{Large $R(z)$ Limit}
  In the strict limit of large $R(z)$ take only leading order of the limit as $R(z)\rightarrow\infty$ of \eqref{eq:vertStaticEom}.
  Taking the leading order term to vanish, we find it is equivalent to
  \begin{equation}
    \label{eq:largeCutoffStaticEom}
    \left(2 f(z)-z f'(z)\right) R'(z)=z f(z) R''(z)\,.
  \end{equation}
  In practice this is equivalent to only considering $\potboundary(R)$.
  The general solution of \eqref{eq:largeCutoffStaticEom} is readily found to be
  \begin{equation}
    \label{eq:largeRsolution}
    R(z) = \constant_1 \int_0^z\frac{y^2 dy}{f(y)}+\rzz \,.
  \end{equation}
  where $\constant_1\in \mathbb R$.
  For the AdS $f(z)=1$, $$R(z) = \rzz  + \frac {\rtz } 6 z^3\,.$$
  For the AdS black brane $f(z)=1 - z^3/\horizonradius^3$, $$R(z) = \rzz -\rtz \frac{\horizonradius^3} {6} \ln\left(1 - z^3/\horizonradius^3\right)\,.$$
  To satisfy the Neumann boundary condition at $z = \horizonradius$ for either geometry for large $\rzz$, $\rtz$ must vanish.
  This implies that any non-trivial deformations of $R$ is partly induced by effects that scale with $R^{\radiuscutoffscaling}$ where $\radiuscutoffscaling < 0$.

  \paragraph{Vortex String Boundary Scalar Expansion}

  Naively, we can treat the scalar field of a vortex string as a free massless scalar field and perform a near-boundary expansion to extract the boundary scalar operator's VEV, $\vev$, and sourcing term. 
  This approach is valid because the $R$ degree of freedom varies the phase of $\phi$, and the modes corresponding to the phase are massless, as discussed in Section \ref{sec:near_boundary_expansion}. 
  The near boundary purely imaginary homogeneous perturbations will be shown to have the same asymptotic behavior: $z^0$ and $z^3$. 
  With the ansatz in \eqref{eq:vortex_ansatz} and a constant normalization, $\phi_{(0)}$, the expansion can be written as:
  \begin{equation}
    \begin{aligned}
      \phi &\sim \phi_{(0)} e^{i \bsphase_{12}(\rzz )}\left(1+\frac{1}{2} i z^2 R^{(2)}(0) \bsphase_{12}'(\rzz )+\frac{1}{6} i \rtz  z^3 \bsphase_{12}'(\rzz )\right)+O\left(z^4\right)\\
           &\sim \phi_{(0)} e^{i \bsphase_{12}(\rzz )}\left(1+\frac{1}{2} i z^2 R^{(2)}(0) \bsphase_{12}'(\rzz ) + \constant_1 \vev z^3 \bsphase_{12}'(\rzz )\right)+O\left(z^4\right)
    \end{aligned}
  \end{equation}
  where $\bsphase_{12}'(\rzz ) \equiv (d \bsphase_{12}/d \rzz ) = sy/\left(y^2 + \left(x + \rzz \right)^2\right) - y/\left(y^2 + \left(x - \rzz \right)^2\right)$. 
  The normalization is necessary because, in the legitimate boundary action, we need to be able to vary it with respect to a sourcing term, which could be the leading order term of $\bscalar$. 
  This leading order term is of constant order and can be identified with $\bscalar$ evaluated at $z=0$, i.e., $\bscalar(0) = \bscalar_{(0)} e^{i\bsphase_{12}}$.

  For a free bulk massless scalar field, the boundary action takes the form $S_{\text{boundary}} \propto \int_\boundary \left( { \bscalar }^{(0)} \right)^2 { \bscalar }^{(3)}$, where $\bscalar \sim { \bscalar }^{(0)} (1 + { \bscalar }^{(3)} z^3 )$. 
  Via the AdS/CFT correspondence, we can identify the source term as ${ \bscalar }^{(0)}$ and the VEV as $\vev = {\delta S_{\text{boundary}}}/{\delta { \bscalar }^{(0)}}$.
  Thus, the third leading order coefficient is identified as the VEV, with $\vev$ being the VEV of the associated boundary scalar operator and $\constant_1$ being an arbitrary constant.

  For a pair of vortex strings, the VEV can be written as
  \begin{equation}
    \label{eq:r3vev}
    \vev \propto \rtz \bsphase_{12}'(\rzz ) = \rtz  \left( \frac{sy}{\left(y^2 + \left(x + \rzz \right)^2\right)} - \frac{y}{\left(y^2 + \left(x - \rzz \right)^2\right)} \right)\,.
  \end{equation}
  While $\vev$ might seem related, it is fundamentally different from the vacuum expectation value (VEV) of a vortex. 
  Unlike a vortex VEV, $\vev$ does not share two key properties: first, it has a non-zero at the exact location where it should be zero for a vortex, and second, it does not exhibit symmetry breaking behavior far away from the vortex (where the VEV typically has a non-zero value).
  Moreover, $\vev$ vanishes for large $r$ and is more like an inverted vortex VEV, as the nonzero values of $\vev$ are localized.

  In the full UV expansion with the UV cutoff, $\ircutoff$, the large energy from the vortices indicates that the back reaction of the geometry must be accounted for. 
  This, coupled with the fact that $\vev$ is localized, provides evidence that such holographic global vortices can induce a phase transition, with the vortices themselves potentially acting as nucleation sites. 
  This phase transition is similar to a cosmological phase bubbles \cite{Grana:2021zvf, Mazumdar:2018dfl}.

  \subsection{Numerical Results}
  \label{sec:numerical_results}
\begin{figure}[htbp]
    \centering
    \includegraphics[width=0.7\textwidth]{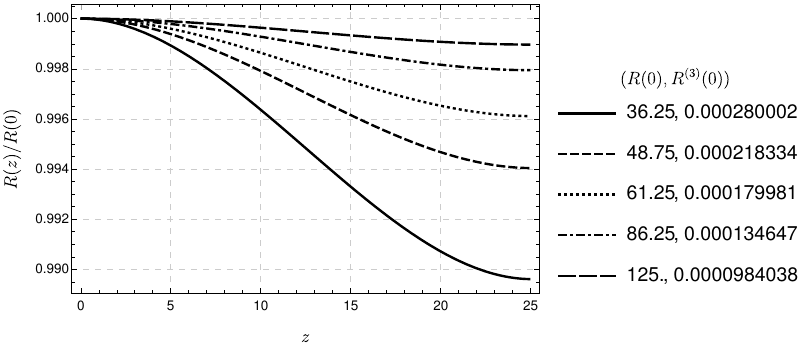}
    \caption{
      For black brane + HW, this is a graph of sample radial profiles, $R(z)/\rzz$ vs $z$.
      $\uvcutoff = 1/1000$, $\radiuscutoff = 10000$, and $\hardwallradius = 25$.
      Across the $x=y=0$ axis, there is a mirror image of $R$ that describes the radial profile of the other vortex string.
    }
    \label{fig:vortex_radial_prof}
  \end{figure}

  In this section, we present the numerical results derived from the vortex approximation discussed in Section \ref{sec:vortexapproximation}. 
  We have two sets of numerical data sets.
  The first set of data is from the shooting method where $\horizonradius$ and $\rzz$ are chosen to fix $R(z)$ and $\rtz$.
  The shooting method is described in Appendix \ref{app:shooting}.
  From this data we derived two sets of plots.
  In the first set of plots we plot $\rtz$ vs $\rzz$.
  For the second set of plots are $U(2\rzz )$ vs $\rzz$ where $U(2\rzz )$ is \eqref{eq:A0vortexpotetialintz}.
  For the reader to better understand found radial profiles, we have displayed a subset of them.
  Figure \ref{fig:vortex_radial_prof} shows a sample profiles of $R(z)$ vs $z$ for a AdS plus hard wall.
  One can see in Figure \ref{fig:vortex_radial_prof} that as $\rzz$ increases the radial profile levels out.
  This is in agreement with the large $R(z)$ solutions \eqref{eq:largeRsolution} where $R(z)$ flattens out for $\rzz\to\infty$.
  Overall radial profiles do not differ noticeably between the spacetimes so, the black brane case will be plotted for these two sets of plots but for the later set.
  For the next numerical results, we present data to analyze the behavior of $\horizonradius$ versus $\rzz$ and $\rtz$.
  The data used for this is several values of $\rtz$ and a range of $\rzz$ values.
  With this input data we solved for $\horizonradius$ and $R(z)$, solving for $\horizonradius$ such that $R^{(1)}(\horizonradius) = 0$.

\begin{figure}[htbp]
    \centering
    \begin{minipage}{.45\textwidth}
      \centering
      \includegraphics[width=\linewidth]{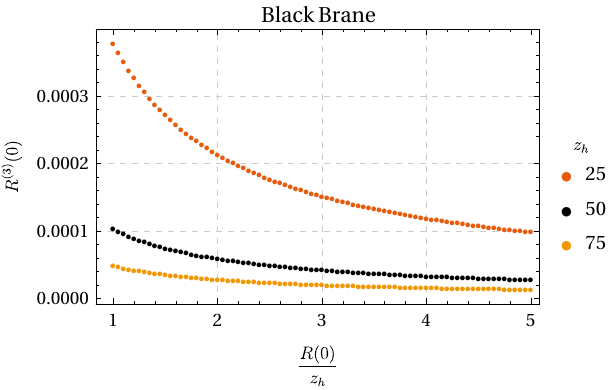}
    \end{minipage}%
    \hspace{0.05\textwidth}
    \begin{minipage}{.45\textwidth}
      \centering
      \includegraphics[width=\linewidth]{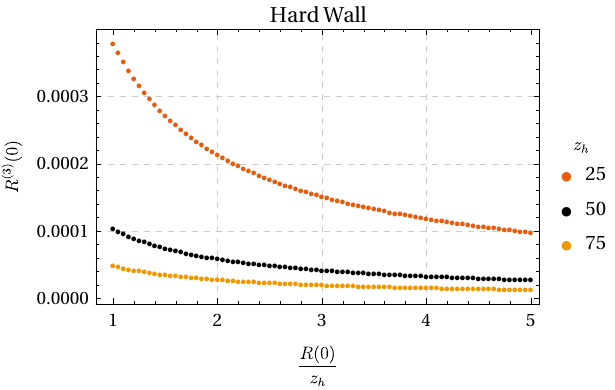}
    \end{minipage}
    \caption{
      For both backgrounds, these are two plots of $\rtz$ vs $\rzz$.
      Black brane is to the left, and AdS plus HW is to the right.
      $\uvcutoff = 1/1000$ and $\radiuscutoff = 10000$ is used.
    }
    \label{fig:r3_vs_r0}
  \end{figure}

  \paragraph{$\rtz$ vs $\rzz$}
  This is the first set of data and plots base off solutions to the boundary value problem defined with \eqref{eq:vertStaticEom} and \eqref{eq:vertStaticBC}, using the shooting method (explained in Appendix \ref{app:shooting}).
  We find solutions for a range of values for $\horizonradius \in \{25, 50, 75\}$ and $\rzz /\horizonradius \in \{1, {(1 + 1/20)} , ..., 5\}$.
  Given these two quantities, we determine the third quantity, $\rtz$, via the aforementioned shooting method.
  For both geometries, the plots of $\rtz$ vs $\rzz$ are displayed in Figure \ref{fig:r3_vs_r0}.
  There are a few notable features of this data.
  First, $\rtz$ decreases for large $\rzz$.
  This bending seems to come from the interaction between these two vortices.
  The farther apart the vortex strings the lower the $\rtz$.
  This also has the possible effect of lowering $\vev$ contributed by both vortex strings.
  Second, it is not shown but for small $\rzz$, there are larger values of $\rtz$ up to a maximum around $\rzz \sim 3/10 \horizonradius$.
  For smaller $\rzz$, the numerical methods used by Mathematica break down, and no more solutions can be found.
  But, our numerical routines fail to converge so $R^{(1)}(\horizonradius) = 0$.
  Because of this the minimum probed value of $\rzz$ is $\sim\horizonradius$.
  Around these values of $R(0) \approx \horizonradius$ we suspect that the vortex string approximation breaks down.

\begin{figure}[htbp]
    \centering
    \begin{minipage}{.45\textwidth}
      \centering
      \includegraphics[width=\linewidth]{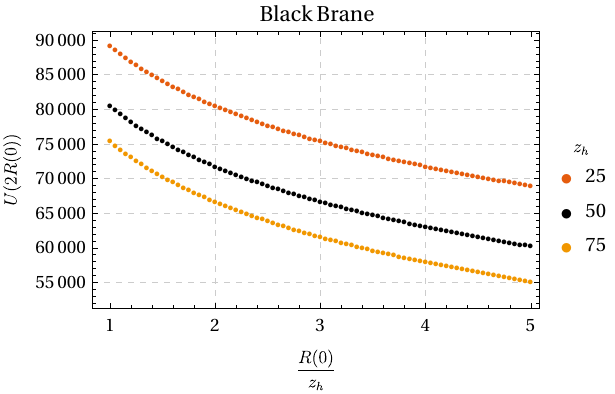}
    \end{minipage}%
    \hspace{0.05\textwidth}
    \begin{minipage}{.45\textwidth}
      \centering
      \includegraphics[width=\linewidth]{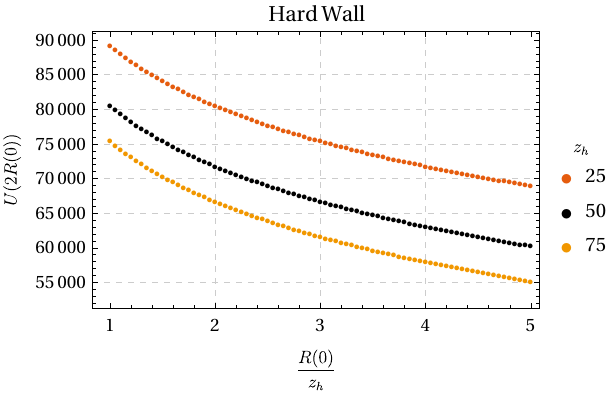}
    \end{minipage}
    \caption{
      For both backgrounds, these are two plots of $\interactionenergy$ vs $\rzz$.
      Black brane is to the left, and AdS plus HW is to the right.
      $\uvcutoff = 1/1000$ and $\radiuscutoff = 10000$ is used.
      Here the vortex-vortex configuration ($s=1$) is taken.
      This potential results in an repulsive force.
      For the vortex-antivortex configuration ($s=-1$), the resulting potentials are just the negated version of above, $-\interactionenergy$.
      This potential results in an attractive force.
    }
    \label{fig:pot_vs_r0}
  \end{figure}

  \paragraph{Potential vs $\rzz$}
  Probing the same values of $\horizonradius$ and $\rzz$, this paragraph examines the potential ($U(2\rzz )$) of the vortex string configurations vs their separation at the boundary.
  For the vortex-vortex configuration, $s=1$, the potential represents a repelling force.
  On the other hand, for the vortex-antivortex configuration, $s=-1$, the potential represents an attractive force.
  These findings agree with what is known about vortex strings \cite{Manton:2004tk}.
  The numerical values over several $\rzz$ can be seen in Figure \ref{fig:pot_vs_r0}.
  A striking feature of the plots in Figure \ref{fig:pot_vs_r0} and Figure \ref{fig:r3_vs_r0} is that the results, up to numerical error, are very similar across geometries.
  This implies that the black brane case for studying probe vortices is sufficient to understand the physics.
  Any back-reaction would then have to factor in the geometry.
  This is in line with the fact that the scalar vacuum is the same for both geometries.

\begin{figure}[htbp]
    \centering
    \includegraphics[width=0.7\linewidth]{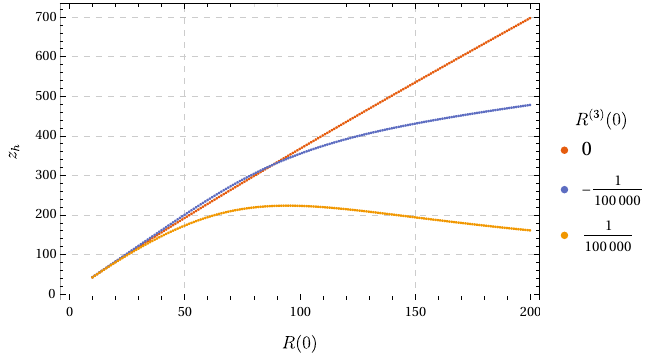}
    \caption{
      For the $\rtz  \sim 0$ case, this is a graph of $z_h$ for several values of $\rzz$ and three values of $\rtz$.
      The geometry is AdS + HW.
    }
    \label{fig:zh_vs_r0_r30zero}
  \end{figure}

  \paragraph{$\horizonradius$ vs $\rzz$}
  The goal is to investigate the behavior of $\horizonradius$ for various values of $\rzz$ and $\rtz$.
  To achieve this, we solve for radius profiles for several values of $\rzz$ and $\rtz$.
  For a chosen $\rzz$ and $\rtz$, we search for $\horizonradius$ such that $R^{(1)}(\horizonradius) = 0$. 
  There are three notable parameter regions.
  \begin{enumerate}
    \item For $\rtz  > \rho_+$, where $\rho_+$ is a small positive number, no termination $z$ value was found.
      The radial profiles diverge by increasing monotonically, and the minimum value of such profiles is the initial value, $\rzz$.
    \item For $\rtz  < -\rho_-$, where $\rho_-$ is a small positive number, the profiles fall towards the center line.
      The larger the absolute value of $\rtz$, the closer the profile ends to the boundary in the bulk.
    \item For $-\rho_- < \rtz  < \rho_+$, we observe the behavior of both solutions, with a minimum at $\horizonradius$.
      For $\rtz  = 0$, $\horizonradius$ is maximized for some $\rzz$ greater than a certain threshold.
      The previous radial profiles solved for with the shooting method belong in this category.
  \end{enumerate}
  \begin{figure}[htbp]
    \centering
    \includegraphics[width=0.7\linewidth]{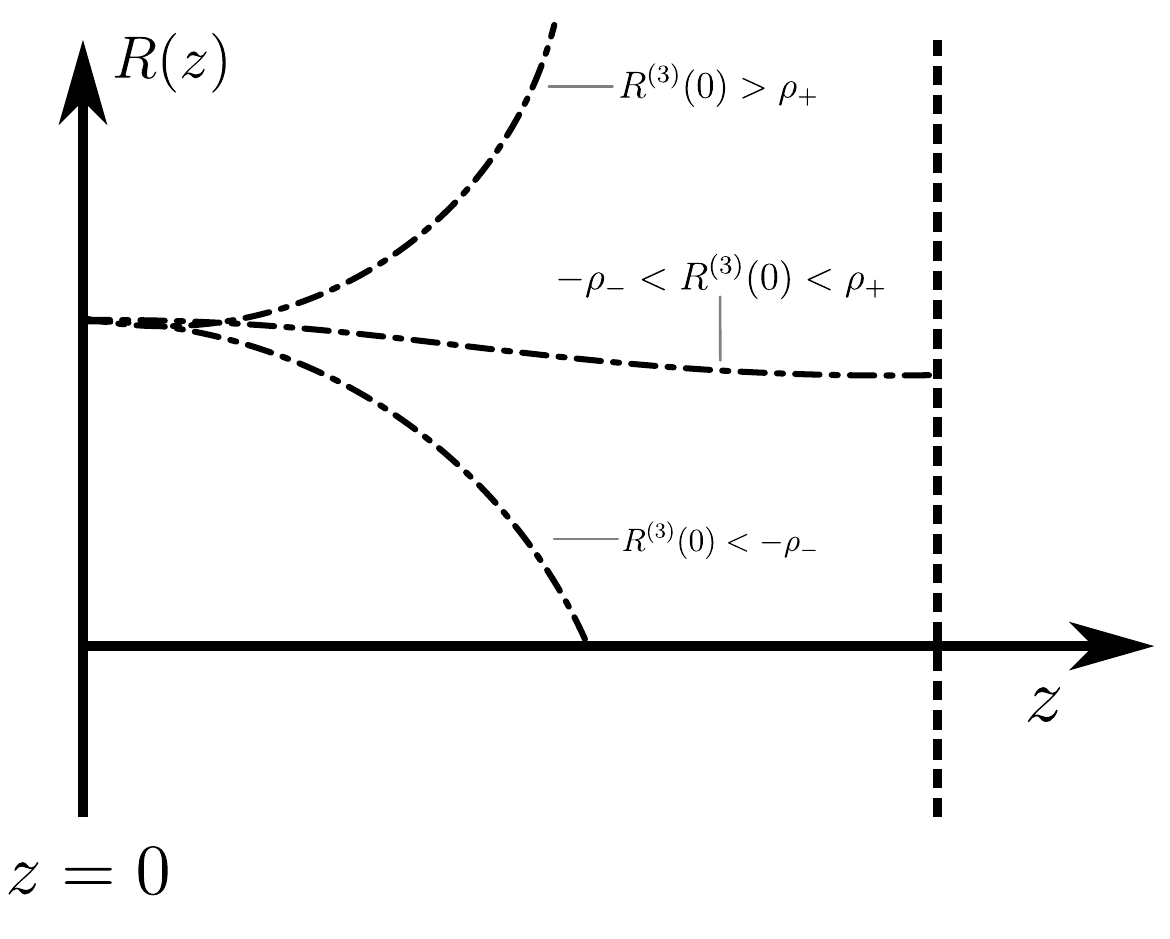}
    \caption{
      A figure that includes a sketch of the three cases of radial profiles.
      For the first two cases the radial profiles do not end on the brane/hard wall.
      Only the middle path of case three hits the the other side and can end on a Neumann boundary condition.
    }
    \label{fig:radial_profile_categories}
  \end{figure}
  Figure \ref{fig:radial_profile_categories} illustrates the three categories.
  For the third case, we have included a plot for several values of $\rtz$ and $\rzz  \in \{10, 11, \ldots, 200\}$ in Figure \ref{fig:zh_vs_r0_r30zero}.
  With similar reasons as the previous two paragraphs, $\horizonradius = 0$ was found for $\rzz$ regardless of $\rtz$.
  We again suspect that the vortex string approximation breaks down here.
  There are a few interesting observations about this numerical data.
  First, for a given $\rzz$, there is a maximum $\horizonradius$.
  In the black brane case, there is a temperature that is inversely proportional to $\horizonradius$, $\temperature\propto1/\horizonradius$.
  Such a maximum $z_h$ in the allowed $\rzz$ implies there is a minimum temperature.
  For a large temperatures, vortices can only get so far before their internal structure becomes relevant.
  The close they get more they bend until they can no longer satisfy the Neumann boundary conditions.
  On the other hand, the allowed region for a given $\rtz$ is to the right of the curves in Figure \ref{fig:zh_vs_r0_r30zero}.
  We have verified this by scanning more values of $\rtz$ for $-\rho_- < \rtz  < \rho_+$.
  One can see that if the $\rtz  = 0$ line continues up for all $\rzz$, the minimum temperature also decreases.
  If this trend were to continue for all $\rzz$ then for $\rzz\to\infty$, $\temperature_\mathrm{min}\to0$.
  This agrees with the derived large $\rzz$ solutions.

  \section{Conclusion}
  \label{sec:conclusion}

  We analyzed a bulk global $\phi^4$ $U(1)$ scalar plus Einstein gravity theory in two $(3+1)$-dimensional asymptotically Anti-de Sitter (AdS) spacetimes: AdS plus hard wall and AdS black brane.
  The bulk scalar field $\bscalar$ is governed by the equation of motion in \eqref{eq:bscalareom} and a quartic potential \eqref{eq:potential}.
  Near the conformal boundary, we found a constant $\bscalar \sim z^0$ solution that preserves the AdS asymptotics and exhibits $U(1)$ symmetry breaking in the bulk, as discussed in Section \ref{sec:near_boundary_expansion}.
  This vacuum is linearly stable against scalar perturbations that do not source boundary operators.
  This novel boundary condition required the scalar mass to be related to the quartic coupling as $\scalarmass^2 = -\quarticcoupling/\adsradius^2$.

  We investigated vortex solutions in the bulk $U(1)$ breaking vacuum, Section \ref{sec:single_vortex}.
  We numerically solved for the magnitude profile, $\magprof$, of a single vortex, centered at $x=y=0$, parallel to the $z$-axis.
  $\magprof$ obeyed a 2nd order non-linear ODE \eqref{eq:singlevortex_mageom}.
  The presence of the the conformal factor $z^2$ prohibited a separation by parts solution.
  Nevertheless, some interesting conclusions can be made if the solution is modeled approximated to be separable.
  We found its width increases with the bulk coordinate $z$, corresponding to a decrease in the effective coupling $\quarticcoupling$.
  Roughly, the conformal factor $z^2$ makes the effect coupling at a cross section of $z$ to be $\sim \quarticcoupling/z^2$, so the radius of the vortex should go like $\sim z/\sqrt \quarticcoupling$.
  This behavior can be seen in Figure \ref{fig:single_vortex_surface_cone_plots}.
  We found that at the boundary, the vortex became singular up to numerical error.
  This necessitated the use of a UV cut off around the singular point on the boundary.
  We believe that this is evidence that the bulk vortex is dual to a nucleation sites for a change of phase on the boundary.

  We also studied vortex string approximation as pairs of interacting vortex-vortex or vortex-antivortex configurations, Section \ref{sec:vortexapproximation}.
  We found that the radial profiles must satisfy Neumann boundary conditions at their termination points, and their unique solution requires specifying the initial location $\rzz$ and the third derivative $\rtz$, Section \ref{sec:boundary_conditions}.
  The global nature of the vortices introduces a log divergence in the energy integral, necessitating UV and IR cutoffs around the boundary ($z\to\infty$) and $r\to\infty$.
  In the limit of large vortex separation, analytical expressions for the radial profiles were found, Section \ref{sec:radialprofileanalysis}.
  Neumann boundary conditions restricted these profiles to only flat profiles.
  This also implied that for pure AdS (without the HW) the radial profile is constant.
  By analogy to a massless scalar field, we found evidence that $\rtz$ proportional to a scalar VEV, $\vev$, Section \ref{sec:radialprofileanalysis}.

  We found vortex string pairs for a given $\rzz$ and $\horizonradius$ and output $\rtz$ in Section \ref{sec:numerical_results}.
  Results confirmed the repulsive and attractive behaviors of vortex-vortex and vortex-antivortex pairs, respectively.
  Further, we found that for large distances the induced candidate $\vev$ decreases with inter-vortex string distance.
  We also found vortex string pairs with a given $\rzz$ and $\rtz$ where $\horizonradius$ is the output.
  For large enough $\left|\rtz \right|$ we found that the string profile diverge from each other for $\rtz  > 0$ and converge to the center ($x=y=0$) for $\rtz  < 0$.
  We also found a maximum $\horizonradius$ for small $\left|\rtz \right|$ corresponding to a minimum temperature $\temperature_\mathrm{min} \propto 1/\rzz$ in the black brane case.

  This study found that the UV cutoff limited what could be said about the boundary theory.
  Nevertheless, in order to integrate to the boundary with vortices their diverging self-energies must be accounted for as back-reaction to the geometry.
  This new boundary geometry would likely differ from a conformally flat geometry.
  One possible connection is that vortices are conformal defects, studied in \cite{Dias:2013bwa}.
  It would interesting to see if the vortex string solutions found in this work could be nucleation sites in a phase transition.
  Furthermore, the vortex string approximation breaks down for small enough $\rzz$, requiring future work where the full vortex solution needs to be found.
  The recent study by Chaffey et al. \cite{Chaffey:2023xmz} explored a similar setup.
  Their method of breaking $U(1)$ symmetry is with a UV brane that has a quartic potential.

  \begin{acknowledgments}
    The work of MA is supported by the \textit{Japan Society for the Promotion of Science Postdoctoral Fellowship for Research in Japan (Short-term)}.
    The work of ME is supported in part by 
    JSPS KAKENHI No. JP22H01221 and the WPI program "Sustainability with Knotted Chiral Meta Matter (SKCM$^2$)" at Hiroshima University.
  \end{acknowledgments}

  \appendix

  \section{Equation of motion Expansion}
  \label{app:eom_expansion}

  This is the expansion of the equation of motion for \eqref{eq:vertStaticEom} near the boundary $z\approx 0$.
  {\tiny
    \begin{equation}
      \begin{aligned}
      &\frac{4 \pi s R'(0) \left(\log \left(\frac{1}{4} \left(\frac{\radiuscutoff^2}{\rzz ^2}+1\right)\right)+1\right)}{z^3} \\
      &+\frac{2 \pi s \left(R''(0) \left(\log \left(\frac{1}{4} \left(\frac{\radiuscutoff^2}{\rzz ^2}+1\right)\right)+1\right)+\frac{\radiuscutoff^2 \left(2-3 R'(0)^2\right)}{\rzz  \left(\rzz ^2+\radiuscutoff^2\right)}\right)}{z^2} \\
      &+\frac{2 \pi \radiuscutoff^2 s R'(0) \left(\left(3 \rzz ^2+\radiuscutoff^2\right) \left(R'(0)^2-2\right)-2 \rzz  \left(\rzz ^2+\radiuscutoff^2\right) R''(0)\right)}{\rzz ^2 z \left(\rzz ^2+\radiuscutoff^2\right)^2} \\
      &-\frac{\pi s}{3 \left(\rzz ^3 \left(\rzz ^2+\radiuscutoff^2\right)^3\right)} \biggl(\radiuscutoff^6 \biggl(-f^{(3)}(0) \rzz ^3 (\log (4)-1) R'(0) \\
      &\quad+6 \rzz  R''(0)+2 R'(0)^4-3 R'(0)^2 \left(\rzz  R''(0)+4\right)\biggr) \\
      &\quad+\rzz ^2 \biggl(R'(0) \biggl(f^{(3)}(0) \rzz  \biggl(\radiuscutoff^6 \log \left(\frac{\radiuscutoff^2}{\rzz ^2}+1\right) \\
      &\quad+\rzz ^2 \left(3 \rzz ^2 \radiuscutoff^2+\rzz ^4+3 \radiuscutoff^4\right) \left(\log \left(\frac{1}{4} \left(\frac{\radiuscutoff^2}{\rzz ^2}+1\right)\right)+1\right)\biggr) \\
      &\quad-2 \rtz  \radiuscutoff^2 \left(\rzz ^2+\radiuscutoff^2\right)^2\biggr) \\
      &\quad+\rzz  R^{(4)}(0) \left(\rzz ^2+\radiuscutoff^2\right)^3 \left(\log \left(\frac{1}{4} \left(\frac{\radiuscutoff^2}{\rzz ^2}+1\right)\right)+1\right)\biggr) \\
      &\quad+6 \rzz ^2 \radiuscutoff^4 \left(4 \rzz  R''(0)+R'(0)^4-2 R'(0)^2 \left(\rzz  R''(0)+3\right)\right) \\
      &\quad+3 \rzz ^4 \radiuscutoff^2 \left(6 \rzz  R''(0)+4 R'(0)^4-3 R'(0)^2 \left(\rzz  R''(0)+8\right)\right)\biggr) \\
      &+\frac{\pi s z}{6 \rzz ^4 \left(\rzz ^2+\radiuscutoff^2\right)^4} \biggl(\rzz ^4 \left(\rzz ^2+\radiuscutoff^2\right)^4 \left(4 f^{(3)}(0) R''(0)+R^{(5)}(0)\right) \log \left(\frac{4 \rzz ^2}{\rzz ^2+\radiuscutoff^2}\right) \\
      &\quad-\left(\rzz ^{12} \left(4 f^{(3)}(0) R''(0)+R^{(5)}(0)\right)\right)+2 \radiuscutoff^2 \biggl(f^{(3)}(0) \rzz ^3 \biggl(3 \left(\rzz ^2+\radiuscutoff^2\right)^3 R'(0)^2 \\
      &\quad-2 \rzz  \radiuscutoff^6 R''(0)\biggr)+R'(0) \biggl(-3 \rzz ^2 \left(\rzz ^2+\radiuscutoff^2\right)^2 \left(3 \rzz ^2+\radiuscutoff^2\right) R''(0)^2 \\
      &\quad-12 \left(4 \rzz ^2 \radiuscutoff^4+5 \rzz ^4 \radiuscutoff^2+10 \rzz ^6+\radiuscutoff^6\right) R'(0)^2 \\
      &\quad+2 \rzz  \left(\rzz ^2+\radiuscutoff^2\right) \left(3 \rzz ^2 \radiuscutoff^2+6 \rzz ^4+\radiuscutoff^4\right) \left(R'(0)^2+6\right) R''(0)\biggr)\biggr) \\
      &\quad-2 \rzz ^4 \radiuscutoff^6 \biggl(\rzz  \biggl(4 \left(2 f^{(3)}(0) \rzz -3 \rtz \right) R''(0)+2 \rzz  R^{(5)}(0) \\
      &\quad-9 R^{(4)}(0) R'(0)\biggr)+5 \rtz  \left(3 R'(0)^2+2\right)\biggr) \\
      &\quad-2 \rzz ^6 \radiuscutoff^4 \biggl(3 \rzz  \biggl(4 \left(f^{(3)}(0) \rzz -\rtz \right) R''(0)+\rzz  R^{(5)}(0) \\
      &\quad-3 R^{(4)}(0) R'(0)\biggr)+7 \rtz  \left(3 R'(0)^2+2\right)\biggr) \\
      &\quad-2 \rzz ^8 \radiuscutoff^2 \biggl(\rzz  \biggl(-4 \left(\rtz -2 f^{(3)}(0) \rzz \right) R''(0)+2 \rzz  R^{(5)}(0) \\
      &\quad-3 R^{(4)}(0) R'(0)\biggr)+3 \rtz  \left(3 R'(0)^2+2\right)\biggr) \\
      &\quad-\rzz ^2 \radiuscutoff^8 \biggl(\rzz  \left(\rzz  R^{(5)}(0)-6 R^{(4)}(0) R'(0)\right) \\
      &\quad+\rtz  \left(-8 \rzz  R''(0)+6 R'(0)^2+4\right)\biggr)\biggr) \\
      &+O\left(z^2\right)
      \end{aligned}
    \end{equation}
  }
  Assuming that $R'(0) = 0$, we get the following simplified expansion.
  \begin{equation}
    \begin{aligned}
      &\frac{2 \pi s \left(R''(0) \left(\log \left(\frac{1}{4} \left(\frac{\radiuscutoff^2}{\rzz ^2}+1\right)\right)+1\right)+\frac{2 \radiuscutoff^2}{\rzz  \radiuscutoff^2+\rzz ^3}\right)}{z^2} \\
      &+\frac{1}{3} \pi s \biggl(R^{(4)}(0) \left(-\log \left(\frac{\radiuscutoff^2}{\rzz ^2}+1\right)-1+\log (4)\right) \\
      &\quad-\frac{6 \left(3 \rzz ^2 \radiuscutoff^2+\radiuscutoff^4\right) R''(0)}{\rzz ^2 \left(\rzz ^2+\radiuscutoff^2\right)^2}\biggr) \\
      &+\frac{1}{6} \pi s z \biggl(4 f^{(3)}(0) (\log (4)-1) R''(0)-\left(4 f^{(3)}(0) R''(0)+R^{(5)}(0)\right) \log \left(\frac{\radiuscutoff^2}{\rzz ^2}+1\right) \\
      &\quad+R^{(5)}(0) (\log (4)-1)+\frac{4 \rtz  \radiuscutoff^2 \left(2 \rzz  \left(\rzz ^2+\radiuscutoff^2\right) R''(0)-3 \rzz ^2-\radiuscutoff^2\right)}{\rzz ^2 \left(\rzz ^2+\radiuscutoff^2\right)^2}\biggr) \\
      &+O\left(z^2\right)
    \end{aligned}
  \end{equation}

\section{Shooting Method}
\label{app:shooting}

\begin{figure}
  \centering
  \includegraphics[width=0.5\textwidth]{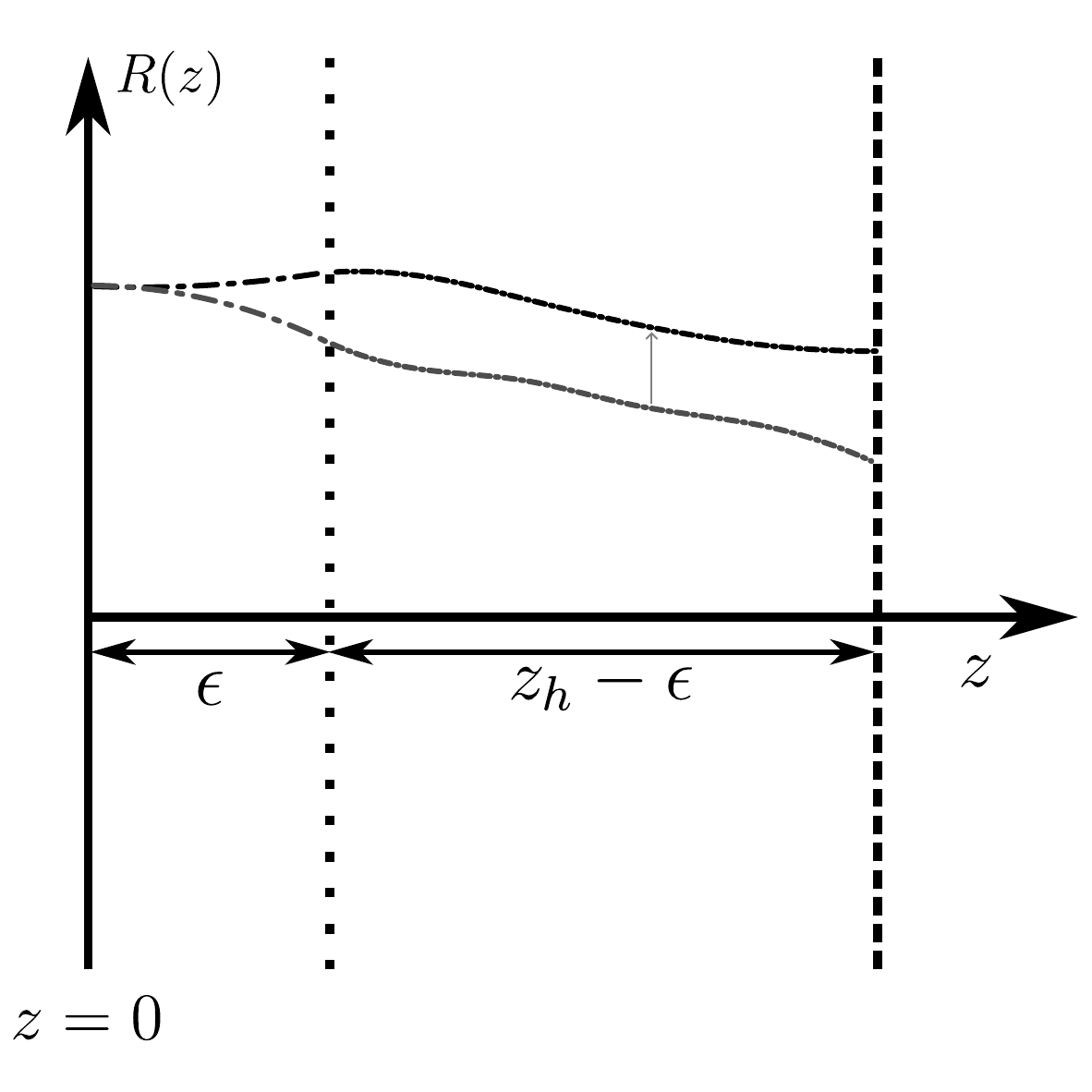}
  \caption{This figure is a sketch of the shooting method. The gray line is an initial guess with fixed $\rzz$ and $\rtz$. These values determine the near-boundary expansion for $0<z<\epsilon$. For $\epsilon < z < z_h$, $R(z)$ can be solved as an initial value problem with initial conditions $R(\epsilon)$ and $R^{(1)}(\epsilon)$. The previous steps implicitly define a function of $\rtz$ (with $\rzz$ and $\horizonradius$ fixed) that outputs $R^{(1)}(\horizonradius)$. A root-finding algorithm can be used to find the root of this function.}
  \label{fig:shooting}
\end{figure}

In this section, we explain the shooting method employed to determine the radial profiles. 
For further clarification, a graphic representation of the shooting method is provided in Figure \ref{fig:shooting}.
Our focus is on the second-order ordinary differential equation \eqref{eq:vertStaticEom}, with an independent variable $z$ and a dependent variable $R(z)$. 
The dependent variable $R(z)$ is subject to Neumann boundary conditions \eqref{eq:vertStaticBC} at both ends of the $z$ domain, $\left(0, z_h\right)$.

We tackle this problem via the shooting method starting at $z=0$. 
For our boundary value problem, we have a singularity at $z=0$, so we can only evaluate the equation near the boundary where $z=\uvcutoff\sim0$. 
The shooting method requires $R(\epsilon)$ and $R'(\epsilon)$. 
We vary the combination of $(R(\epsilon),R'(\epsilon))$ such that the Neumann boundary condition at $z=z_h$ is satisfied.
However, our desired parameters are $\rzz$ and $\rtz$. 
To obtain these, we rewrite $\left(R(\epsilon),R'(\epsilon)\right)$ in terms of $\left(\rzz ,\rtz \right)$  near-boundary expansion given as
\begin{equation}
  \begin{aligned}
    R(\epsilon) &\approx \rzz +\frac{\radiuscutoff^2 \epsilon^2}{\rzz  \left(\rzz ^2+\radiuscutoff^2\right) \left(-\log \left(\frac{\radiuscutoff^2}{4\rzz ^2}+\frac 14\right)-1\right)}+\frac{1}{6} \rtz  \epsilon^3 + \ldots\,,\\
    R'(\epsilon) &\approx \frac{2\radiuscutoff^2 \epsilon}{\rzz  \left(\rzz ^2+\radiuscutoff^2\right) \left(-\log \left(\frac{\radiuscutoff^2}{4\rzz ^2}+\frac 14\right)-1\right)}+\frac{1}{2} \rtz  \epsilon^2 + \ldots\,.
  \end{aligned}
\end{equation}
up to fifth order for our case.

With this expansion of $R$ near the boundary, we can evaluate $R(\uvcutoff)$ and $R^{(1)}(\uvcutoff)$ to be the initial conditions of an initial value problem. 
From $z=\uvcutoff$, $R$ can be numerically integrated to $z = \horizonradius$.
With $\rzz$ fixed, we can use $\rtz$ as a shooting parameter, varying such that Neumann boundary condition at $z=\horizonradius$ is satisfied.
The steps of the shooting method can be summarized as follows:
\begin{enumerate}
  \item Develop a near-boundary expansion around $z=0$, parameterizing it with variables $\rzz$ and $\rtz$.
  \item With the near-boundary expansion, evaluate $R(\epsilon)$ and $R^{(1)}(\epsilon)$.
  \item Using $R(\epsilon)$ and $R^{(1)}(\epsilon)$, solve the equation of motion from $z=\epsilon$ to the other end of the domain at ${z = \horizonradius > 0}$.
  \item Employ a numerical solver to adjust the initial values $\rzz$ and $\rtz$ such that $R^{(1)}(z_h)$ is zero.
\end{enumerate}

\bibliographystyle{jhep}
\bibliography{ms}

\end{document}